\documentclass[slac_one]{revtex4}
\usepackage{graphicx}
\usepackage{fancyhdr}
\usepackage{multirow}
\begin{document}

\title{Search for Isolated Leptons with Missing {\boldmath $p_T$} and Multi-leptons at HERA} 

\author{M. Turcato (on behalf of the H1 and ZEUS Collaborations)}
\affiliation{Universit\"at Hamburg, Institut f\"ur Experimentalphysik, Luruper Chaussee 149, 22761 Hamburg, Germany}
\begin{abstract}
A search for events with one or more isolated leptons in the final state
is performed on a data sample collected in $e^\pm p$ collisions with the
H1 and ZEUS detectors at the HERA collider. The data sample corresponds to
an integrated luminosity of $\sim 1~{\rm fb}^{-1}$, representing the full
HERA high-energy data set. The yields of single, di- and tri-lepton
events are measured and compared to the Standard Model predictions,
looking for possible deviations. No significant discrepancy with respect
to the Standard Model expectations is observed.
\end{abstract}

\maketitle

\thispagestyle{fancy}

\section{INTRODUCTION} 

At the electron-proton collider, HERA, which operated at a center-of-mass energy of $\sim 319~{\rm GeV}$, possible signatures of physics beyond the Standard Model (SM) are investigated by looking for events in which one or more high transverse momentum ($p_T$) leptons are found in the final state.
The leptons provide a clean event signature, and the investigation of the high mass, high $p_T$ regions, where the SM expectation is low, could reveal some signal of new physics. 

The two experiments ZEUS and H1 at HERA are extensively studying such events, using the full available statistics. The search is focused on topologies in which one isolated lepton (electron or muon) is found in the final state together with missing $p_T$ ($p_T^{\rm miss})$, and on topologies in which more than one isolated lepton is found.
\section{ISOLATED LEPTONS IN EVENTS WITH MISSING TRANSVERSE MOMENTUM}
In the SM, the dominant contribution to the production of a high-$p_T$ isolated lepton together with missing $p_T$ is the production of real $W$ bosons, where the $W$ decays into a lepton and a neutrino. The SM cross section of this process, known at next-to-leading order (NLO) with an uncertainty of 15\%~\cite{wnlo}, is of the order of $\sim 1~{\rm pb}$.  The event simulation is based on the leading-order {\sc Epvec} Monte Carlo (MC) generator, reweighted to the $W$ cross section at NLO~\cite{wnlo2}. The main background sources in this search come from the misreconstruction of the leptons or of the energy in the event.  Neutral current deep inelastic scattering (NC-DIS) events, $ep\rightarrow eX$, in which the energy is not properly reconstructed, can lead to fake $p_T^{\rm miss}$ and therefore to a fake signal in the electron channel; charged current events, which have genuine $p_T^{\rm miss}$, can mimic a signal if a fake lepton is identified in the event; dimuon events, in which one of the two muons is not identifed, can also mimic a fake signal in the muon channel.  Alternative event samples were used to verify that the fake signal rates were well simulated by the MC. 

The H1 Collaboration analysed the HERA I data sample (luminosity, ${\cal L} = 118.3~{\rm pb}^{-1}$) and reported~\cite{h1_isol_pub} an excess of this kind of events with respect to the SM predictions. The analysis was then extended to the full data sample collected by H1~\cite{h1_isol_prel}, corresponding to an integrated luminosity of $478~{\rm pb}^{-1}$. 
The sample consists of $184~{\rm pb}^{-1}$ of data collected in electron-proton collisions, and $294~{\rm pb}^{-1}$ of data collected in positron-proton collisions.  The event selection required a  lepton (electron or muon) with $p_T^l> 10~{\rm GeV}$ in the angular range $5^\circ < \theta^l< 140^\circ$. The lepton had to be isolated from the rest of the event: the isolation was quantified using the distance between the lepton and the nearest jet or the nearest track in the $\eta-\phi$ plane. The event  had also to show a large imbalance in transverse momentum,  reconstructed using the calorimeter and the tracking information. 
\begin{table}[!ht]
  \centering
  \begin{tabular}{|c|c|c|c|c|}
    \hline
    \multirow{2}{*}{\textbf{Data sample} }& \multicolumn{2}{|c|}{\textbf {Full sample}} & \multicolumn{2}{|c|}{\boldmath{$p_T^X>25~{\rm GeV}$}} \\ \cline{2-5}
                                 & \textbf{Data} & \textbf{SM} & \textbf{Data} & \textbf{SM}\\
    \hline
    $e^+p$ (294 pb$^{-1}$) & 41 & $34.5\pm4.8$ & 21 &   $8.9\pm1.5$\\
    $e^-p$ (184 pb$^{-1}$) & 18 &  $24.4\pm3.4$ &  3 &    $6.9\pm1.0$\\
    $e^\pm p$ (478 pb$^{-1}$) & 59 &  $58.9\pm8.2$ &  24 &    $15.8\pm2.5$\\
    \hline
  \end{tabular}
  \caption{H1 results of the search for events with an isolated high-$p_T$ lepton and $p_T^{\rm miss}$. The data are compared with the SM predictions, separately for  $e^+p$, $e^-p$ and for the full sample, combining the electron and the muon channels. }
  \label{tab:h1_isol}
\end{table}

 The results obtained by the H1 Collaboration are shown in Table~\ref{tab:h1_isol}. The number of observed events is shown separately for the $e^+p$ and the $e^-p$ data samples, as well as for the full data sample, for the whole $p_T^X$ region and for a subsample at large $p_T^X$, where $p_T^X$ is the transverse momentum of the hadronic system. The overall agreement between the data and the SM expectations, which are dominated by $W$ production, is good, but some excess is observed for $p_T^X>25~{\rm GeV}$. 
The excess is restricted to the $e^+ p$ data sample: limiting the analysis to this sample, the discrepancy between the data and the Monte Carlo predictions is of the order of $3\sigma$.
 
 A similar excess was not confirmed by the ZEUS Collaboration in a recent publication~\cite{zeus:W} using the full HERA data sample (${\cal L}=504~{\rm pb}^{-1}$). This analysis used the same event selection as of the H1 Collaboration, but the angular region of the lepton was restricted to $15^\circ<\theta^l < 120^\circ$ in order to ensure a good detector acceptance.  The ZEUS Collaboration reported good agreement between the data and the SM predictions for both the full sample and the high-$p_T^X$ region. For $p_T^X>25~{\rm GeV}$, 6 events were observed in the positron sample compared to an expectation of $7.4\pm1.0$, and  11 events were found in the full data sample, to be compared with an expectation of $12.9\pm1.7$.

\begin{figure}[!ht]
  \centering
  \includegraphics[width=0.45\columnwidth]{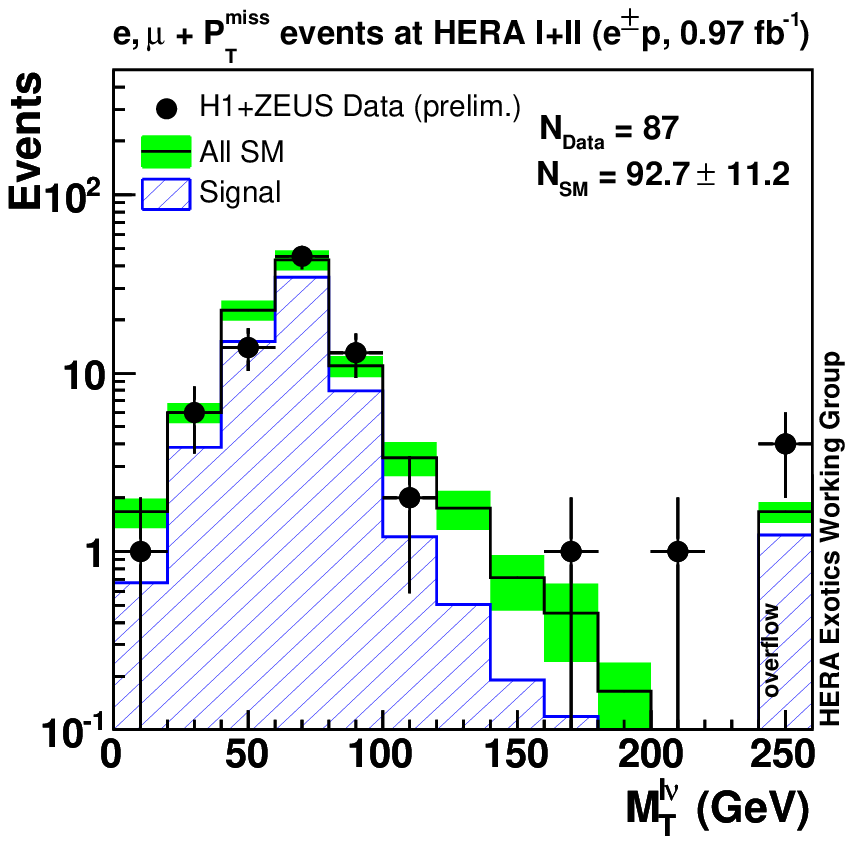}
  \includegraphics[width=0.45\columnwidth]{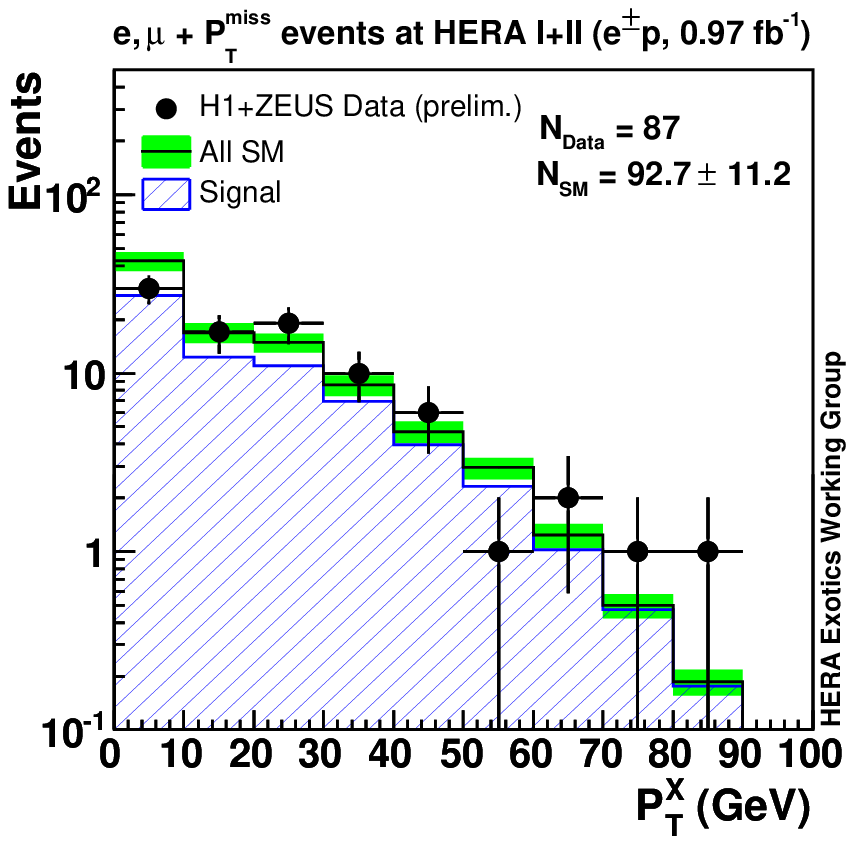}
  \caption{Transverse mass, $M_T^{l\nu}$, (left) and hadronic transverse momentum, $p_T^X$, (right) distributions for the combined H1+ZEUS analysis of events with an isolated lepton and missing $p_T$ in the final state. The data (points) are compared to the SM expectations (open histogram) which are dominated by real $W$ production (hatched histogram). The shaded band shows the total uncertainty on the SM expectation.}
  \label{fig:h1zeus_isol}
\end{figure}

The analyses from the two experiments were compared: using the {\sc Epvec} Monte Carlo it was found that the two detectors had compatible efficiency in the kinematic region where they were directly comparable. Most of the high-$p_T^X$ events observed by H1 fell into the region in which the two analyses overlapped. A common phase space was therefore defined by restricting the H1 analysis to the ZEUS angular region.
The results of the combined analysis~\cite{h1zeus_isol} are shown in Fig.~\ref{fig:h1zeus_isol} and in Table~\ref{tab:h1zeus_isol}.  Figure~\ref{fig:h1zeus_isol} shows the transverse mass ($~M_T^{l\nu} = \sqrt{2p_T^l p_T^\nu(1-\cos \phi^{l\nu})}~$) and the hadronic transverse momentum distributions of the combined data sample. The $M_T^{l\nu}$ distribution is compatible with the Jacobian peak expected for $W$ production. Similarly, the $p_T^X$ distribution is compatible with the SM expectations, peaking at low values for $W$ production. A hint of an excess in the $e^+p$ data at high $p_T^X$ remains, where 23 events were observed compared to $14.6\pm 1.9$ expected from the SM.%
\begin{table}[!ht]
  \centering
  \begin{tabular}{|c|c|c|c|c|}
    \hline
    \multirow{2}{*}{\textbf{Data sample} }& \multicolumn{2}{|c|}{\textbf {Full sample}} & \multicolumn{2}{|c|}{\boldmath{$p_T^X>25~{\rm GeV}$}} \\ \cline{2-5}
                                 & \textbf{Data} & \textbf{SM} & \textbf{Data} & \textbf{SM}\\
    \hline
    $e^+p$ (0.58 fb$^{-1}$) & 57 & $53.1\pm6.4$ & 23 &   $14.6\pm1.9$\\
    $e^-p$ (0.39 fb$^{-1}$) & 30 &  $39.6\pm5.0$ &  6 &    $10.6\pm1.4$\\
    $e^\pm p$ (0.97 fb$^{-1}$) & 87 &  $92.7\pm11.2$ &  29 &    $25.3\pm3.2$\\
    \hline
  \end{tabular}
  \caption{ZEUS and H1 combined results of the search for events with an isolated high-$p_T$ lepton and $p_T^{\rm miss}$. The data are compared with the SM predictions, separately for  $e^-p$, $e^+p$ and for the full sample, combining the electron and the muon channels. }
  \label{tab:h1zeus_isol}
\end{table}
\section{MULTI-LEPTON EVENTS}
 
 In the SM, the production of multi-lepton events in $ep$ collisions proceeds mainly via photon-photon interactions~\cite{vermaseren}. Since this is a quantum electrodynamic (QED) process, the cross section is precisely calculable in the SM. Multi-lepton events are simulated using the {\sc Grape}~\cite{abe} MC program, which includes all  electroweak matrix elements at tree level. The dominant source of background for multi-electron production arises from NC-DIS events. QED Compton scattering, $ep \to e\gamma X$, also contributes. 

The ZEUS Collaboration presented new preliminary results~\cite{zeus:dimuon} on the production of high-$p_T$ muon pairs, based on the full HERA data (${\cal L}=444~{\rm pb}^{-1}$). The analysis was done inclusively, asking at least two muons in the event, with $p_T>5~{\rm GeV}$, in the angular region $20^\circ<\theta^\mu<160^\circ$. 
 Differential cross sections as a function of the di-muon invariant mass, of the transverse momentum  of the highest $p_T$ muon, and of the sum of the $p_T$'s of the two muons were measured and compared to the {\sc Grape} MC. Good agreement between the data and the SM predictions was found, also in the high-mass and high-$p_T$ regions.

 The H1 and ZEUS collaborations~\cite{h1prelim:07-166} have  measured multi-electron production combining the full HERA data sample collected by the two experiments, corresponding to an integrated luminosity of 0.94~fb$^{-1}$.
In order to coherently combine the results of both experiments a common phase space was established.
 The final multi-electron selection required that there were at least two central ($20^\circ < \theta < 150^\circ$) electron candidates, of which one had to have $p_T > 10$~GeV and the other $p_T > 5$~GeV. Additional electrons were searched for in the angular range $5^\circ < \theta < 175^\circ$.

\begin{figure}[!ht]
  \centering
  \includegraphics[width=0.38\columnwidth]{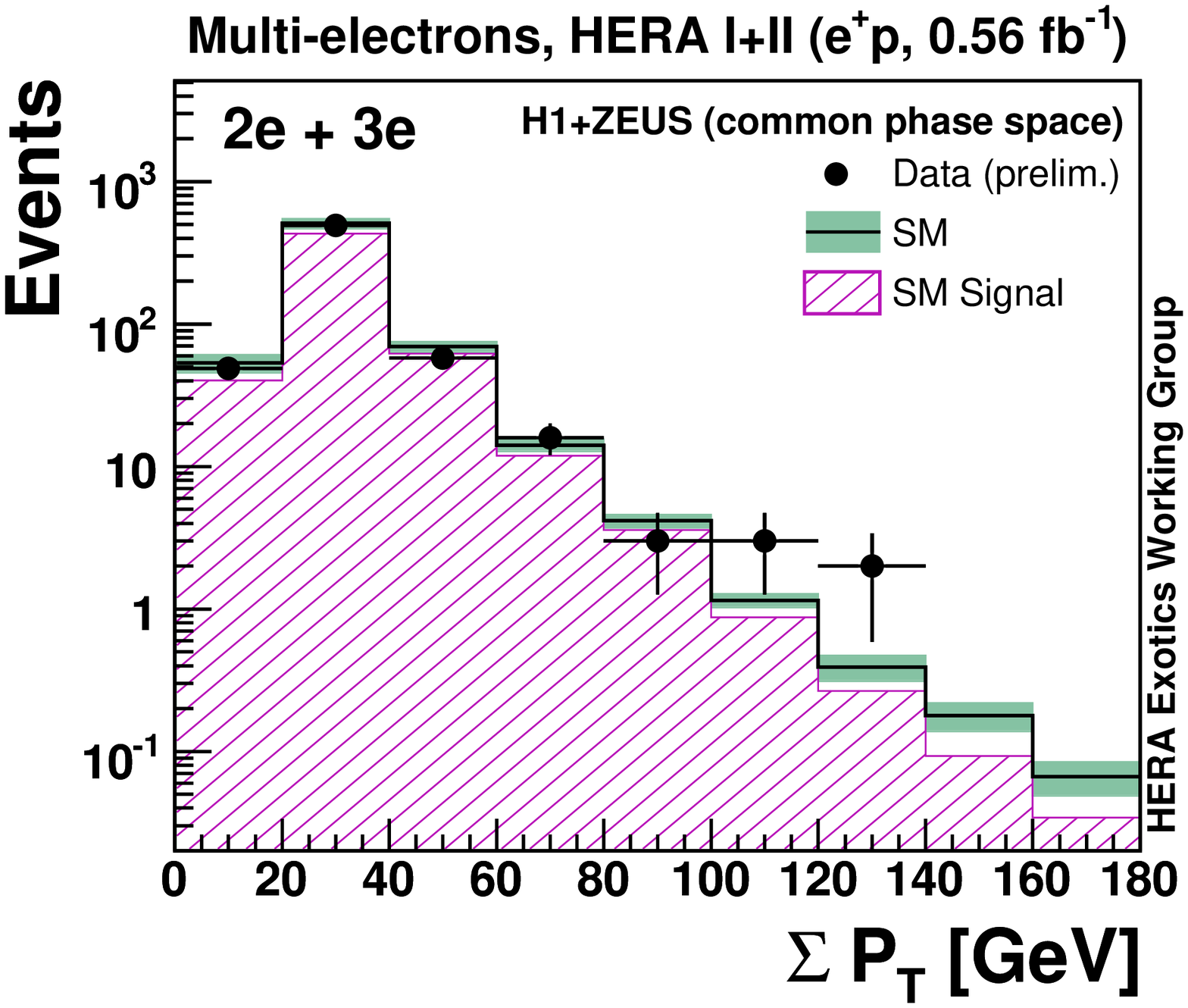}
  \includegraphics[width=0.38\columnwidth]{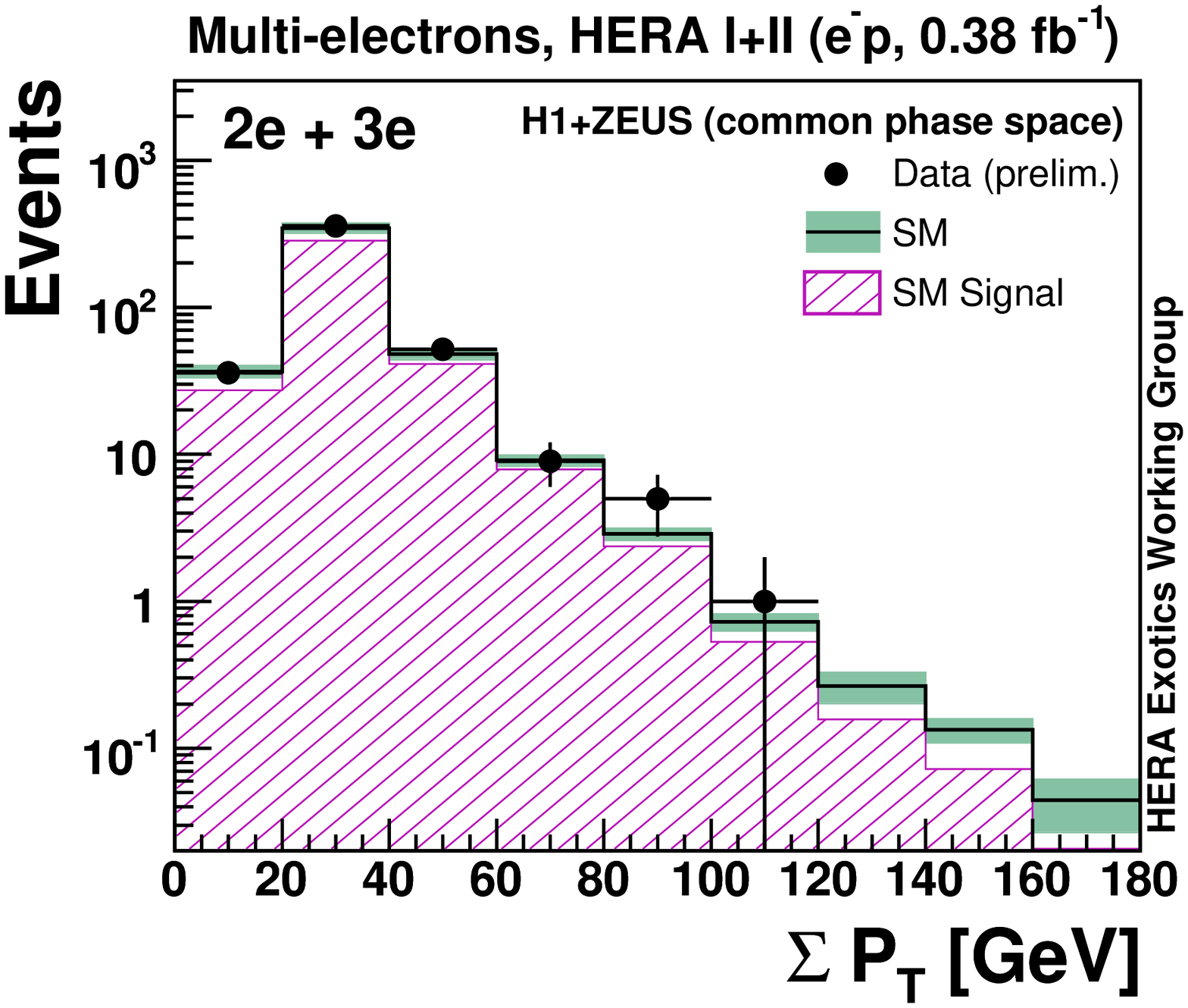}
  \caption{Distributions of the scalar sum of the transverse momenta of the combination of 2 and 3 electrons events compared to expectations for data taken in $e^+p$ (left) and $e^-p$ (right) collisions.}
  \label{fig:diele1}
\end{figure}

 The observed event yields are in good agreement with SM expectations, which are dominated by lepton pair production. The distribution of the scalar sum of all electron's $p_T$ ($\sum p_T$) is shown in Fig.~\ref{fig:diele1}.

When requiring $\sum P_T > 100$~GeV, 6 events are left, whilst $3.0\pm0.3$ are expected from the SM.
In $e^+p$ collisions 5 events are observed, whilst $1.8\pm0.2$ are expected from the SM (see Table~\ref{tab:diele2}); only one event is found in $e^-p$ data. 
\begin{table}[!ht]
  \centering
  \begin{tabular}{|c|c|c|c|c|}
    \hline
    \textbf{Data sample} & \textbf{Data} & \textbf{SM} & \textbf{Pair Production} & \textbf{NC-DIS + Compton}\\
    \hline
    $e^+p$ (0.56 fb$^{-1}$) & 5 & $1.82\pm0.21$ & $1.28\pm0.16$ &
      $0.54\pm0.10$\\
    $e^-p$ (0.38 fb$^{-1}$) & 1 & $1.19\pm0.14$ & $0.79\pm0.09$ &
      $0.40\pm0.08$\\
     All (0.94 fb$^{-1}$)  & 6 & $3.00\pm0.34$ & $2.07\pm0.24$ &
      $0.94\pm0.16$\\
    \hline
  \end{tabular}
  \caption{Yield of events with $\sum P_T > 100$~GeV for H1 and ZEUS data combined. The errors on the prediction include model uncertainties and experimental systematic errors added in quadrature.}
  \label{tab:diele2}
\end{table}

The H1 Collaboration recently published~\cite{h1_multilep} the final multi-lepton analysis based on the full HERA data sample, ${\cal L}=463~{\rm pb^{-1}}$. All the possible topologies in which muons or electrons are found in the final state were exclusively considered. Therefore, new event topologies such as 
 $e\mu$, $e\mu\mu$, $ee\mu$, $eeee$ were analysed, together with the di-muon, di- and tri-electron final states already studied. 
 The final multi-lepton selection required that there were at least two central ($20^\circ < \theta < 150^\circ$) lepton (electron or muon) candidates, one of which had to have $p_T > 10$~GeV and the other $p_T >$ 5~GeV. Additional electrons were searched for in the angular range $5^\circ < \theta < 175^\circ$, and additional muons in $20^\circ < \theta < 160^\circ$. According to the flavours of the identified leptons, these samples were classified into the different topologies listed above. The observed event yields are in good agreement with SM expectations, which are dominated by pair production.
When requiring $\sum P_T > 100$~GeV, 5 events are observed in $e^+p$ collisons, whilst $1.6\pm0.2$ are expected from the SM;
 no events are observed at high $\sum P_T$ in $e^-p$ data.

\section{CONCLUSIONS}
The ZEUS and H1 Collaborations at the electron-proton collider HERA perform searches for events with an isolated high-$p_T$ lepton and missing $p_T$, and for multi-lepton events. The cross sections of these processes are well known in the Standard Model and deviations in the high mass and high $p_T$ region could reveal a signal of new physics. All the observations are in good agreement with the SM predictions, although a few high-$p_T$ events are observed in the $e^+p$ data in a region where the expectations from the SM are low.

\end{document}